\documentclass[runningheads]{llncs}
\usepackage[T1]{fontenc}
\usepackage{hyperref}
\usepackage{amsmath}
\usepackage{amsmath}  
\usepackage{amsfonts} 
\usepackage{caption}
\usepackage{graphicx}
\usepackage{subcaption}
\usepackage{tcolorbox}
\usepackage{xcolor}
\usepackage[export]{adjustbox}
\usepackage{soul}
\begin{document}
\title{One Shot GANs for Long Tail Problem in Skin Lesion Dataset using novel content space assessment metric}
%
%
\author
{Kunal Deo\inst{1}\orcidID{0009-0004-4409-5571}\and 
Deval Mehta\inst{2}\orcidID{0000-0002-6907-7589} \and
Kshitij Jadhav\inst{1}\orcidID{0000-0001-9795-8335}}
%
%
\institute{Koita Centre for Digital Health, Indian Institute of Technology, Bombay \\
\and
AIM for Health Lab, Department of Data Science \& AI, Faculty of IT, Monash University \\
\email{deo.kunal.a@gmail.com, deval.mehta@monash.edu, kshitij.jadhav@iitb.ac.in}}

\maketitle              

\begin{abstract}
Long tail problems frequently arise in the medical field, particularly due to the scarcity of medical data for rare conditions. This scarcity often leads to models overfitting on such limited samples. Consequently, when training models on datasets with heavily skewed classes, where the number of samples varies significantly— a problem emerges. Training on such imbalanced datasets can result in selective detection, where a model accurately identifies images belonging to the majority classes but disregards those from minority classes. This causes the model to lack generalizability, preventing its use on newer data. This poses a significant challenge in developing image detection and diagnosis models for medical image datasets.
To address this challenge, the One Shot GANs model was employed to augment the tail class of HAM10000 dataset by generating additional samples. Furthermore, to enhance accuracy, a novel metric tailored to suit One Shot GANs was utilized.

\end{abstract}
\vspace{-2.5em}
\section{Introduction}

Deep learning has been widely utilized in the medical field for classification of medical images belonging to different classes \cite{Deep_Learning_Medicine}. 
The first deep-learning models were used to aid in the diagnosis of breast
cancer using a dataset consisting of large number of Mammograms \cite{Deep_Learning_Mammogram}. This
dataset brought to light a problem of long tail misclassification that
has persisted in many image classification problems. The long tail of
medical data is especially problematic since it could result in the
misdiagnosis resulting in downstream problems.

With the rapid rise of computer-aided diagnostics, it is becoming ever
so important that trained models make accurate classifications on a
dataset. This problem especially presents itself when a set of very
skewed classes is brought into the picture. This results in the model
misclassifying the skewed classes because of underrepresentation resulting in fall in the overall accuracy of the model. Over the years multiple solutions have been presented to deal with problems resulting from long tail distributions such as class re-balancing, information augmentation and module
improvement \cite{Long_tail_survey}. Class re-balancing aims to adjust the distribution of classes in the training dataset, often through oversampling minority classes or undersampling majority classes. Information augmentation leverages techniques to artificially increase the size of underrepresented classes in the dataset, typically through data augmentation methods that create additional, synthetic examples. Module improvement involves enhancing the deep learning models themselves, making them more robust to imbalances in the data and enhancing the learning of minority classes.
\par
The proposed solution in this paper leverages One-shot Generative Adversarial Networks (GANs) \cite{One_Shot_GANs}, which is a novel method at an intersection between data augmentation and class re-balancing in which samples of an underrepresented class are generated using a single training image. This innovative approach addresses the scarcity of data in underrepresented classes by generating high-quality synthetic examples, thereby providing a richer and more balanced dataset for training deep learning models. By incorporating One-shot GANs into the training process, we aim to mitigate the impact of long-tail distributions on model performance, enhancing the model's ability to make accurate and reliable classifications across all classes, which is crucial in the high-stakes domain of medical diagnosis.


\section{Our Contribution}
\vspace{-2em}
\begin{figure}[h]
  \centering
  \begin{tcolorbox}[colframe=black, boxrule=1pt, colback=white] 
    \includegraphics[width=\textwidth]{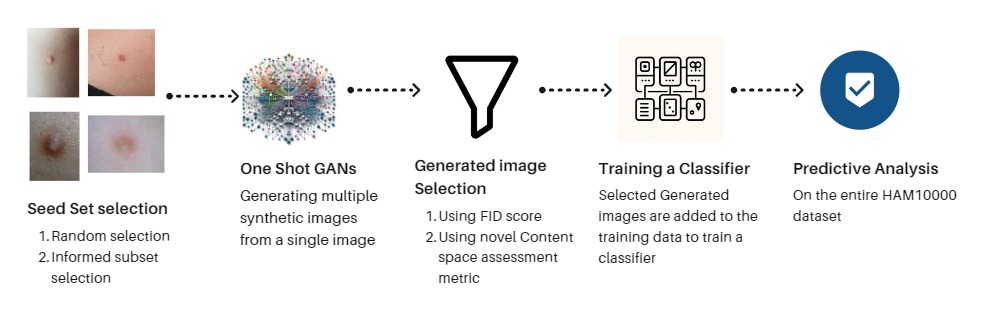}
  \end{tcolorbox}
  \caption{\textbf{Flowchart for the process of the proposed solution.}}
\label{fig: Process flowchart}
\end{figure}
\vspace{-2em}
\begin{itemize}
    \item We have used a method for efficient image selection (informed subset selection) to select the seed set of images for data augmentation using one-shot GANs. We observed that this helped boost the ultimate accuracy of the classifier on the minority class.
    \item We have utilized One-Shot-GANs for class re-balancing for minority classes which suffer from extreme skewness. We have also compared this to WGANs and observe a significant boost of accuracy on minority class after utilizing One-Shot-GANs 
    \item We have devised a novel metric that is called the \textbf{content-space assessment} to select generated images post data augmentation using one-shot GANs. This metric increases the accuracy of the classifier when trained on the dataset post-augmentation compared to when the same is done with FID scores.    
\end{itemize}

\section{Preliminaries}

\subsection{Subset Selection of Images}
We optimized the selection of images from the training set to compensate
for the similarity of the images generated using One-Shot GANs. This was
done to prevent the overfitting on the training dataset
post-augmentation and result in the highest possible accuracy. Using the
Submodlib library the most diverse images were selected from the
Dermatofibroma class. \cite{Submodlib}

The Disparity Sum function (Equation \ref{eqn_submod}) was utilized to find the most diverse images of the chosen class. The Disparity Sum function calculates the sum of the pairwise distance and uses this to model the diversity of a given image \cite{Submodlib}.
\vspace{-0.5em}
\begin{align}
f_{DSum}(X)=\sum_{i,j \in X} d_{ij} \label{eqn_submod} 
\end{align}

Equation \ref{eqn_submod} above is used to calculate the sum of pairwise distances
between all the elements within a subset. This Disparity Sum function is utilized for this purpose as it has the capacity to include
outliers in a subset if it boosts its diversity. This helps maximize the
diversity of a given subset making it ideal for selecting the most diverse images possible.

\subsection{Wasserstein GAN (WGANs)}
To compare the solution that we have designed we have utilized WGANs to create a baseline model as this is a very common model which is used for generating images as it is very effective at avoiding the problem of mode collapse. The basic model of a GANs consists of a generator G and a discriminator D. The Generator utilizes noise input to generate images after learning from a training dataset. Following this, a Discriminator acts as the Generator's 'adversary' and tries to distinguish between the fake generated images and the real images from a test set. \cite{GANs_Paper} 

Let c be the real samples, G(z) be the fake images generated by generator G, V be the function that calculates the adversarial loss and p(z) be the probability distribution of input noise z. Then the objective of the entire model can be described in equation \ref{eqn_GANs} below.  
\vspace{-0.5em}
\begin{align}
\min_{G} \max_{D} V(D, G) = \mathbb{E}_{x \sim p_{\text{data}}(x)} \left[\log D(x)\right] + \mathbb{E}_{z \sim p_{z}(z)} \left[\log(1 - D(G(z)))\right] \label{eqn_GANs}
\end{align}

The WGANs is a type of Generative Adversarial Network in which the objective is to minimize the Wasserstein or Earth-Mover Distance. This helps in attaining a smoother convergence which makes WGANs a lot more stable than regular GANs. \cite{WGANs}   
\vspace{-0.5em}
\begin{align}
W(p_g,p_r) = \underset{y\in \prod(p_g,p_r)}{inf}E_{(x,x'\sim y)}||x - x'|| \label{eqn_Wasserstein}
\end{align}
where \( \Pi(P_r, P_g) \) is the set of all joint distributions \( \gamma(x, y) \) whose marginals are respectively \( P_r \) and \( P_g \).

\subsection{One-Shot GANs}
One-Shot GANs is a model which utilizes a single image to generate multiples
samples. This is achieved by having a dual branch discriminator that
judges the context and layout of a generated image separately. This is
done to overcome the memorization effect which results in the
overfitting of the GANs model \cite{One_Shot_GANs}.
\vspace{-1em}
\begin{figure}[h]
\begin{tcolorbox}[colframe=black, boxrule=1pt, colback=white] 
\includegraphics[width=\textwidth]{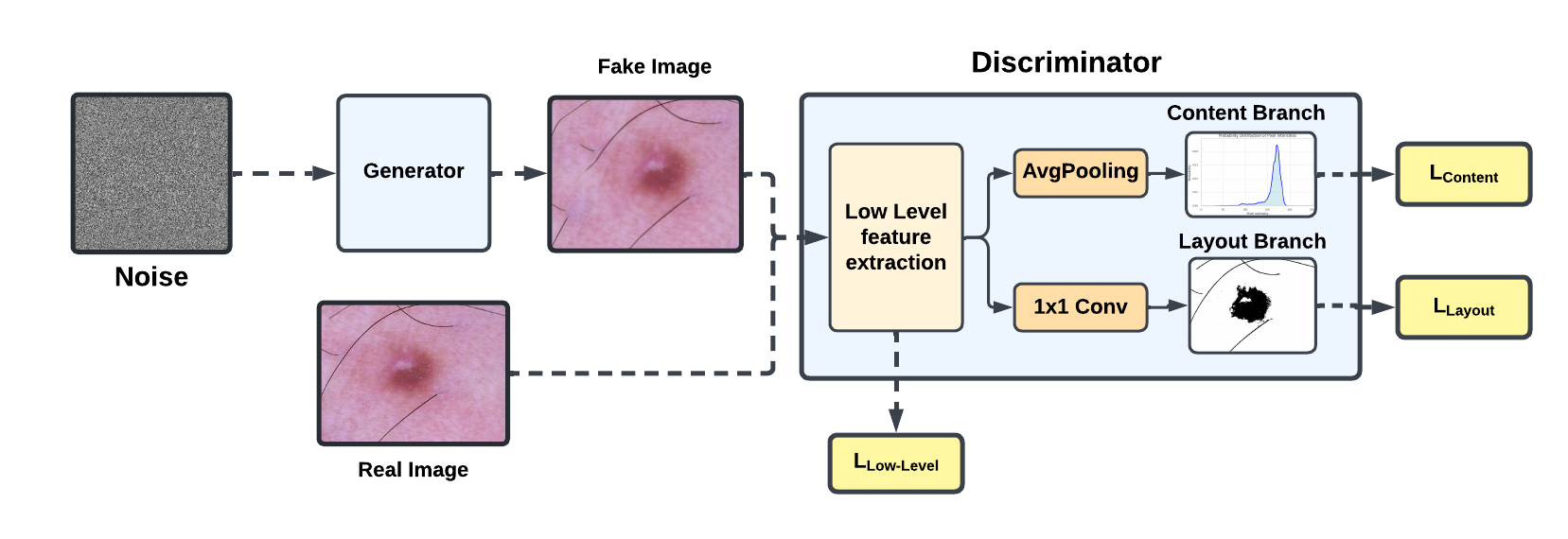}
\end{tcolorbox}
\caption{ One-Shot GAN. The two-branch discriminator judges the content distribution separately from the scene layout realism and thus
enables the generator to produce images with varying content and global
layouts\cite{One_Shot_GANs}.}
\end{figure}
\vspace{-1em}
Low-level features are extracted after which the discriminator is split into two branches. The Content Branch assesses the layout of an image independent of its position in space while the Layout branch assesses the distribution of pixels in an image independent of their intensities. Along with this we also have to consider that One-Shot GANs-generated images needs to generate perceptually different images independent of their latent codes. For the purpose we utilize equation \ref{eqn_LDR} below. \cite{One_Shot_GANs}
\vspace{-1em}
\begin{align}
    L_{DR}(G) = \mathbb{E}_{z_1,z_2} \left[ \frac{1}{L} \sum_{l=1}^{L} \| G_l(z_1) - G_l(z_2) \|_k \right] \label{eqn_LDR} 
\end{align}

Where \(L_{DR}(G)\) is the diversity Regularization loss term which is independent of distance in latent space. Where $\| \mathbf{G}_l(\mathbf{z}) \|_1$ denotes the L1 norm, and $\mathbf{G}_l(\mathbf{z})$ indicates features extracted from the $l$-th block of the generator $G$ given the latent code $\mathbf{z}$. \cite{One_Shot_GANs}

Following this the overall adversarial loss is calculated for each part of the discriminator. The overall adversarial loss of the model is calculated by utilizing the adversarial loss of each branch as seen in equation \ref{eqn_Ladv} . \cite{One_Shot_GANs}
\begin{align}
L_{adv}(G, D) = L_{D_{content}} + L_{D_{layout}} + 2L_{D_{low-level}} 
  \label{eqn_Ladv} 
\end{align}

Following this we calculate the overall objective of the One-Shot GANs utilizing equation \ref{eqn_OneShot} as given below. \cite{One_Shot_GANs}
\begin{align}
\min_{G} \max_{D} \left( L_{adv}(G, D) - \lambda L_{DR}(G) \right)\label{eqn_OneShot} 
\end{align}

where $\lambda$ controls the strength of the diversity regularization and $L_{adv}$ is the adversarial loss from Equation \ref{eqn_Ladv}.


\subsection{Fréchet Inception Distance}

FID score calculates the amount
of dissimilarity between the original and generated images. It is a
popular metric used to find the quality of the images generated by GANs
\cite{FID_paper}.

Let $\mu_{\text{real}}$ and $\mu_{\text{gen}}$ be the means of the activations of real and generated images, $\sigma_{\text{real}}$ and $\sigma_{\text{gen}}$ are the covariance matrices of the activations of real and generated images, respectively.
\vspace{-0.5em}
\begin{align}
FID = \lVert\mu_{\text{real}} - \mu_{\text{gen}}\rVert^2_2 + \text{Tr}(\sigma_{\text{real}} + \sigma_{\text{gen}} - 2(\sigma_{\text{real}}\sigma_{\text{gen}})^{1/2}) \label{eqn_FID}
\end{align}

For this selection process, the FID scores were calculated for the images generated from each seed image. Using these FID scores we select 10 generated images from each seed image. Since there are 10 initial seed images we end up with 100 selected images in total (Look at supplementary section for more details). 


\subsection{ Content-Space assessment}

\begin{figure}
    \centering
    \includegraphics [width=\textwidth]{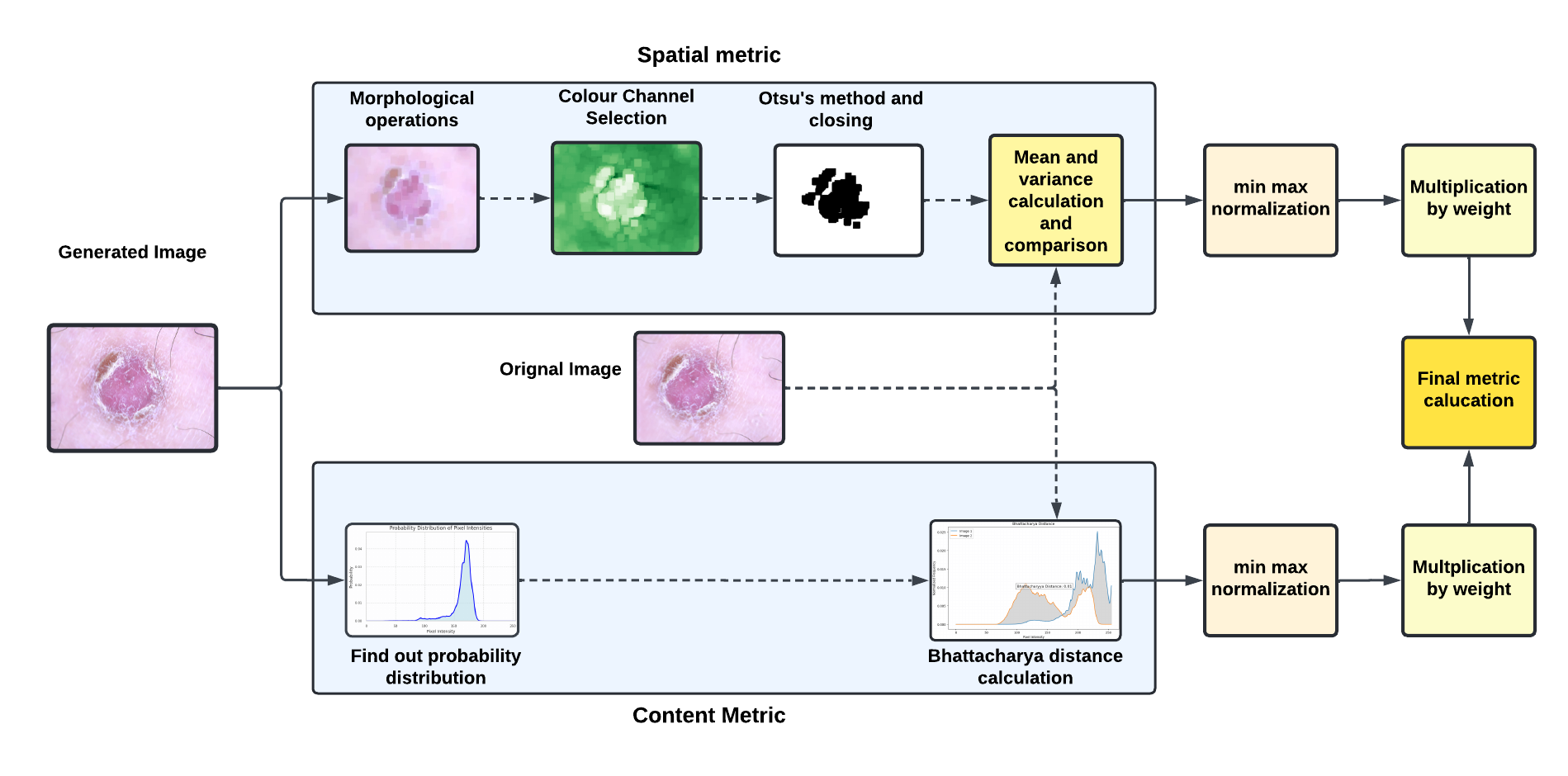}
    \caption{Flowchart describing the process Content-space assessment}
    \label{fig:Context-Space}
\end{figure}

Content-Space assessment is a new metric developed by us that is utilized to select
images generated by One-Shot-GANs. This metric was designed while
keeping in mind the dual branch discriminator that is present in
One-Shot-GANs. The metric uses two probability distributions - the first one is dependent on the distribution of pixel intensity and doesn't consider spatial distribution, whereas the other one considers the spatial distribution of pixels and is independent of their intensities as shown in Figure \ref{fig:Context-Space}.
\vspace{-1.5em}
\subsubsection{Content assessment metric}

The first part of the metric assesses content independent of space. This
is done by calculating the number of pixels according to their intensity
and plotting a probability distribution for them. The probability
distribution for the original image is found along with the probability
distribution for images generated using One-Shot GANs. The probability distributions we obtain are multinomial representations of the pixels within an image, and we use Bhattacharya Distance (Equation \ref{eqn_Bhatt_cal}) as a metric to compare both the images to gauge their dissimilarity to one another. Bhattacharya distance was chosen due to its symmetrical property and its lack of reliance on any prior distribution assumption \cite{Bhattacharya_distance}.
\vspace{-1em}
\subsubsection{Pre-processing for Spatial metric}

\begin{figure}[h]
  \begin{subfigure}{0.24\textwidth}
\includegraphics[width=\textwidth, frame]{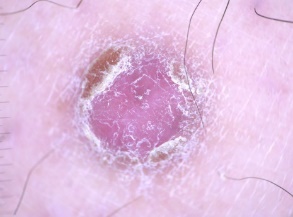}
\caption{$ $}
\label{fig: Morphology a}
  \end{subfigure}
   \begin{subfigure}{0.24\textwidth}
 \includegraphics[width=\textwidth, frame]{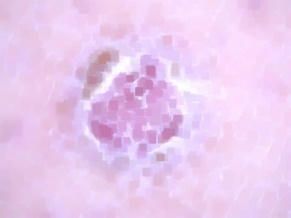}
 \caption{$ $}
 \label{fig: Morphology b}
  \end{subfigure}
   \begin{subfigure}{0.24\textwidth}
\includegraphics[width=\textwidth, frame]{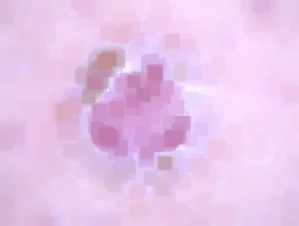}
\caption{$ $}
\label{fig: Morphology c}
  \end{subfigure}
   \begin{subfigure}{0.225\textwidth}
 \includegraphics[width=\textwidth, frame]{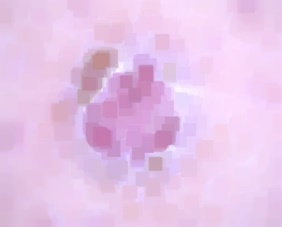}
 \caption{$ $}
 \label{fig: Morphology d}
  \end{subfigure}
  \caption{\textbf{(a)} The original image selected for
Morphological operation \textbf{(b)} Closing operation is performed
\textbf{(c)} Image after Closing and Erosion \textbf{(d)} Image after
Closing, Erosion and Interpolation}
\end{figure}

Each image is passed through multiple morphological operations. This is
done because images have much noise such as hair, variations in
brightness, and sometimes small air bubbles. Preprocessing steps were
used to eliminate this noise and prepare the images for segmentation.
This process consists of performing a closing operation on the image
followed by erosion and interpolation. This results in an image which
contains only the most important features needed for segmentation (Figure \ref{fig: Morphology d})
\cite{auto_segment}.
\vspace{-2em}

\vspace{-3.5em}

\subsubsection{Channel Selection}
\begin{figure}[h]
  \begin{subfigure}{0.24\textwidth}
\includegraphics[width=\textwidth, frame]{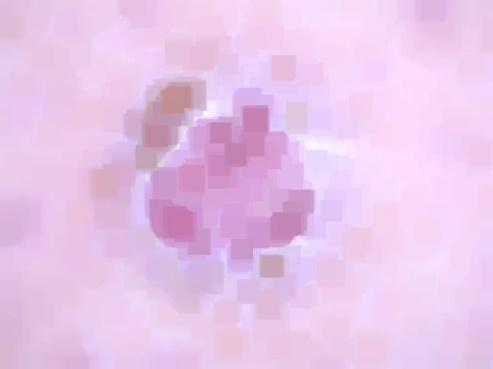}
\caption{$ $}
\label{fig: Channel a}
  \end{subfigure}
   \begin{subfigure}{0.24\textwidth}
 \includegraphics[width=\textwidth, frame]{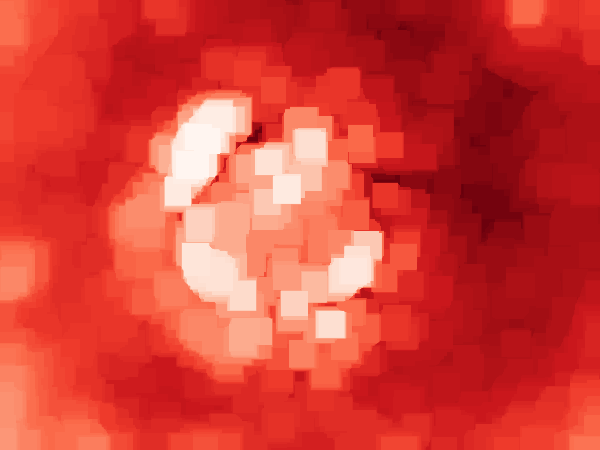}
 \caption{$ $}
 \label{fig: Channel b}
  \end{subfigure}
   \begin{subfigure}{0.24\textwidth}
\includegraphics[width=\textwidth, frame]{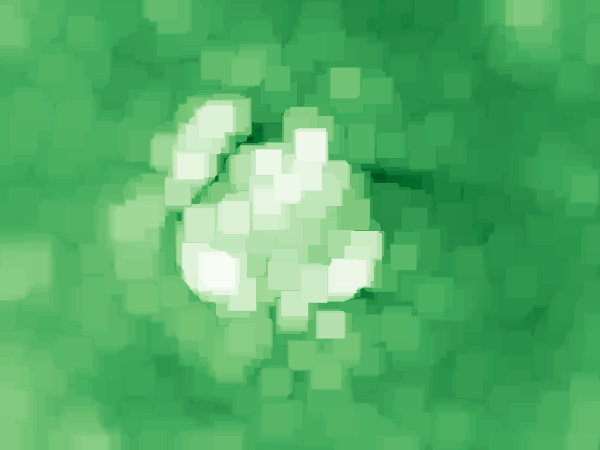}
\caption{$ $}
\label{fig: Channel c}
  \end{subfigure}
   \begin{subfigure}{0.24\textwidth}
 \includegraphics[width=\textwidth, frame]{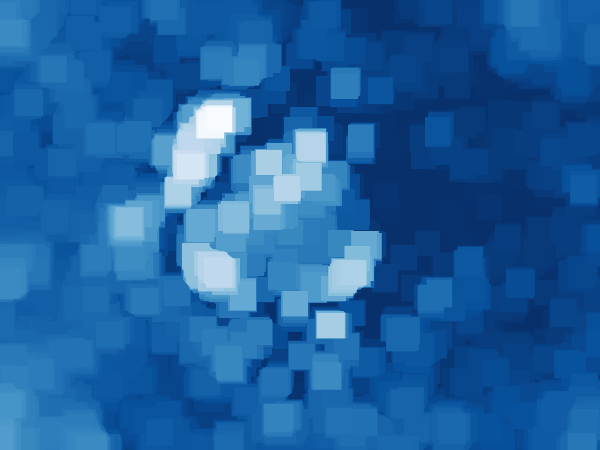}
 \caption{$ $}
\label{fig: Channel d} 
  \end{subfigure}
  \caption{\textbf{(a)} images obtained after morphological
operations \textbf{(b)} Red channel analysis of the Tumor \textbf{(c)}
Green channel analysis of the Tumor \textbf{(d)} Blue channel analysis
of the Tumor. }
\end{figure}

After Morphological operations are carried out which results in the
final image as shown in Figure \ref{fig: Morphology d}, the image is then split into its
component RGB channels which are then extracted separately. We have selected the green channel (Fig \ref{fig: Channel c}) to use for Otsu's method (Refer Supplementary Data Section). \cite{auto_segment}. 
\vspace{-1em}
\subsubsection{ Otsu's method}

After Extracting the RGB color channels separately we need to segment
the skin lesion from the rest of the skin. To perform this task,
Otsu's Method is used to distinguish between background and foreground.
The Otsu's Threshold is calculated for each individual color channel and
then segmentation is performed \cite{otsu}.

\begin{figure}[h] \centering
   \begin{subfigure}{0.24\textwidth}
       \includegraphics[width=\textwidth, frame]{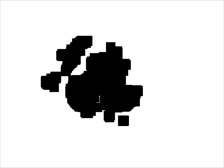}
   \caption{$ $}    
   \label{fig:otsu a}
   \end{subfigure}
    \begin{subfigure} {0.24\textwidth}
        \includegraphics[width=\textwidth, frame]{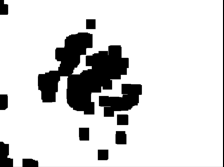} 
    \caption{$ $}
    \label{fig:otsu b}
    \end{subfigure}
    \begin{subfigure} {0.24\textwidth}
        \includegraphics[width=\textwidth, frame]{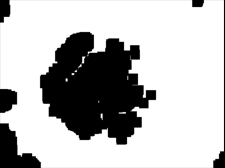}
    \caption{$ $}
    \label{fig:otsu c}
    \end{subfigure}

\caption{\textbf {(a)} Otsu's Thresholding and closing on green channel of image. \textbf{(b)} Otsu's Thresholding and closing on Blue Channel.
\textbf{(c)} Otsu's Thresholding and closing on Red Channel of image }
\vspace{-1.60em}
\end{figure}

As seen from Figures \ref{fig:otsu a}, \ref{fig:otsu b} and \ref{fig:otsu c} , green channel Otsu's thresholding results in the most significant portion of information being captured while avoiding any possible dermatological noise found in the image (Supporting data in Supplementary Section). A final closing operation is performed after Otsu's method is utilized to fill up any gaps that might be present in the final segmentation so we can obtain two distinct components within the image
and no stray clusters are remaining.

After this step the mean and variance of the cluster is calculated to
determine the center point of the cluster along with the spread of
pixels in the segmented area. The mean and variance of this cluster in
generated image is then compared to the mean and variance of the cluster
in the original image.
\vspace{-1em}
\subsubsection{ Selection of images}

In the Content branch the Bhattacharya distance is calculated between
the distributions of the original image and the generated image. The
scores are calculated by first flattening the two-dimensional image
array into a one-dimensional array and the following equation is used to
find out the Bhattacharya distance between the two distributions of
pixels. After all the Bhattacharya distances are calculated for the
images the results are normalized using min-max normalization. 
In Equation \ref{eqn_Bhatt_cal}, \(p_i\) and \(q_i\) are the probabilities associated with the i-th outcome. The summation of these probabilities is taken across the discrete distribution to give the Bhattacharya distance \(D_B\).\cite{Bhattacharya_distance_signal}
\begin{align}
D_B(p,q) = -\ln \left( \sum_{i} p_i q_i \right) \label{eqn_Bhatt_cal}
\end{align}

In the Spatial branch of the metric, we utilize color channel selection
and Otsu's method for segmentation. Using this we obtain the cluster of
pixels that represent the tumor and the background. For the following
operations it is understood that Otsu's method and K-means algorithms
have similar objective functionality. \cite{Otsu_kmeans}

This implies that mathematically, the cluster of pixels that result from
Otsu's can be treated as a single cluster of global k-means algorithm.
The centroid for this cluster is calculated by taking the average of
both x and y coordinates of cluster pixels. The Following Equations \ref{eqn_Centroid+X},\ref{eqn_Centroid_Y} and \ref{eqn_Centroid_mean} below describe the process:-
\vspace{-1em} 
\begin{align}
    X_{\text{Centroid}} &= \frac{1}{n}\sum_{i = 1}^{n}x_{i}    &&\qquad \label{eqn_Centroid+X} \\
    Y_{\text{Centroid}} &= \frac{1}{n}\sum_{i = 1}^{n}y_{i}    &&\qquad \label{eqn_Centroid_Y} \\
    \text{Centroid} &= \frac{X_{\text{Centroid}} + Y_{\text{Centroid}}}{2}   &&\qquad \label{eqn_Centroid_mean}  
\end{align}

Using this cluster of pixels, a boundary box is defined to further
reduce any latent noise left after Otsu's method. The Region of interest
is justified using the longest continuous contour present in the
segmented image. The standard deviation of the ROI is then calculated
which provides us with the spread of pixels around the centroid of the
cluster. Since Otsu's is similar in objective functionality to k-means,
the standard deviation gives us the spread of data from the centroid. The standard deviation is given in Equation \ref{eqn: Standard_Deviation} below: -
\begin{align}
\sigma = \sqrt{\frac{1}{N \times M} \sum_{i=1}^{N} \sum_{j=1}^{M} (x_{ij} - \mu)^2} \label{eqn: Standard_Deviation}
\end{align}
Here N is the number of rows in the image while M is the number of columns. \(x_ij\) is the value of the pixel at position ixj and \(\mu\) is the mean intensity of all pixels. 

Following this the difference between the Standard Deviation and
Centroid of the original image are compared to the generated images. The combination of these differences gives us a Spatial score as seen in Equation \ref{eqn_Spatial_Score} below. Once the
Spatial scores are calculated they are normalized using min-max
normalization.
\begin{align}
 SS = \left| \left( {Centroid}_{original} - {Centroid}_{generated} \right) \right| + \left| \left( \sigma_{original} - \sigma_{generated} \right) \right|\label{eqn_Spatial_Score}
\end{align}
The results of both the context branch and the spatial branch are then
combined by assigning weight and summing the individual scores. A weight
\emph{w1 and w2} are assigned to the Context branch and Spatial Branch
respectively. The weights are assigned with a condition as described in Equation \ref{eqn_condition} below: -
\begin{align}
  {0 \leq w}_{1} \leq 1;0 \leq w_{2} \leq 1 \label{eqn_condition}
\end{align}

Let C be the Bhattacharya distance of an image and S be the spatial score
of the image. Then the final metric can be described with Equation \ref{eqn_Context_Space} below.
\begin{align}
          Content Space Metric=w_1*C+w_2*S \label{eqn_Context_Space}                            
\end{align}

When all the metric scores are calculated a final min-max normalization
is applied to the scores. This score is used to select images generated by One-Shot-GANs
\vspace{-1mm}
\subsection{Classifier utilized} \label{sec:CNN}

A basic 24-layer CNN model was utilized to gauge the effectiveness of all experiments. This CNN model Utilized Convolutional, Batch Normalization, AveragePooling layers which all had a ReLu activation function. For the output layer a Dense layer was defined with softmax activation function and 7 outputs. \cite{kaggle_code}

\section{Methodology and Results}

To test the proposed method in Figure \ref{fig: Process flowchart}, we utilized the HAM10000 Skin
Cancer dataset \cite{HAM10000_dataset}. The classes in this dataset were heavily skewed (Table \ref{table:metadata}) which makes it a perfect example to demonstrate the proposed solution.
The main issue with the One-Shot GANs method for augmentation is the
risk of overfitting on classifiers since samples are generated using a
single training image. To reduce this risk, we emphasis diversity of
selected samples. This is done during the selection of training images
used for One-Shot GANs driven data augmentation. After generation of images, the results are selected by utilizing two different metrics which are compared further in the paper.

\begin{table}
    \centering
    \vspace{-2em}
          \caption{ Class wise distribution of samples in metadata}
    \begin{tabular}{|c|c|} \hline 
         \textbf{Class Name}& \textbf{Number of Samples}\\ \hline 
         Melanocytic nevi (nv)
& 6705\\ \hline 
         Melanoma (mel)
& 1113\\ \hline 
         Benign Keratosis-like lesions (bkl)
& 1099\\ \hline 
         Basal Cell carcinoma (bcc)
& 514\\ \hline 
         Actinic Keratoses (akiec)
& 327\\ \hline 
         Vascular Lesions (vasc)
& 142
\\ \hline 
         Dermatofibroma (df)
& 115
\\ \hline
    \end{tabular}
    
    \label{table:metadata}
\vspace{-2.5em}   
\end{table}


\subsection{Train Test Splitting and Selection of augmentation class}

The data was split into training and test sets. The test set consisted
of 40 samples taken from each class of the 7 classes of the dataset at
random. This amounts to a total of 280 samples taken for the test set.
The remaining 9735 samples were present in the un-augmented Training
set. This method was used to ensure that the same number of samples from each class were present in the test set.


Using this training set a Convolutional Neural Network (refer Section \ref{sec:CNN}) was trained
without any augmentation of the training set. The main problem being
addressed here is the skewed of classes of HAM10000 dataset, the
hypothesis is that lesser number samples in a particular class result in
lower accuracies for that class. The
following class wise accuracies were obtained as shown in Table \ref{tab:Classwise_CNN_preaugmentation} below to test this hypothesis. 
An association can be drawn using Table \ref{tab:Classwise_CNN_preaugmentation} and Table \ref{table:metadata}. The Dermatofibroma class possesses the least number of samples and gives the lowest accuracy of all the classes. To correct this problem, new samples need to be added to the class so the Convolutional Neural Network can be trained to produce a higher accuracy on the class while maintaining overall accuracy of the model.
\vspace{-2em}

\begin{table} [h]
    \centering
\caption{Class wise accuracies of CNN on test set pre-augmentation}
\label{tab:Classwise_CNN_preaugmentation}
    \begin{tabular}{|c|c|} \hline 
         \textbf{Class Name}
& \textbf{Accuracy}
\\ \hline 
         Melanocytic nevi (nv)
& 0.95
\\ \hline 
         Melanoma (mel)
& 0.40
\\ \hline 
         Benign Keratosis-like lesions (bkl)
& 0.475
\\ \hline 
         Basal Cell carcinoma (bcc)
& 0.675
\\ \hline 
         Actinic Keratoses (akiec)
& 0.40
\\ \hline 
         Vascular Lesions (vasc)
& 0.80
\\ \hline 
         Dermatofibroma (df)
& \textbf{0.0}
\\ \hline
 \textbf{Overall Test accuracy}
&0.54
\\\hline
    \end{tabular}

\end{table}
\vspace{-2em}


\vspace{-1.25em}
\subsection{Random Selection and One Shot GANs}  \label{sec:Random_OneShot}

We try to address the class imbalance problem by randomly selecting 10
seed images which are used to train the One-Shot GANs model. The One
Shot GANs produces 100 images for each of these 10 randomly selected
seeds giving 1000 generated samples. A further selection criterion is
utilized to select 100 generated samples which are used to augment the
training set. These 100 generated samples were picked by utilizing FID score metric for selecting 10 generated images from each of the 10 seed images. This was done by selecting images that give the lowest FID scores.

\vspace{-1.5em}
\subsubsection{Results:}
The accuracy of the CNN was found post augmentation of the training
dataset with 100 samples generated by One-Shot GANs. The model used is the same as the one mentioned in Section \ref{sec:CNN}.
\vspace{-1em}
\begin{table}[h]
    \centering
\vspace{-2em}    
\caption{Class wise accuracies of CNN by Random Selection  }
\label{tab:CNN_Random_Selection}
    \begin{tabular}{|c|c|} \hline 
         \textbf{Class Name}
& \textbf{Accuracies}
\\ \hline 
         Melanocytic nevi (nv)
& 0.925
\\ \hline 
         Melanoma (mel)
& 0.375
\\ \hline 
         Benign Keratosis-like lesions (bkl)
& 0.55
\\ \hline 
         Basal Cell carcinoma (bcc)
& 0.625
\\ \hline 
         Actinic Keratoses (akiec)
& 0.45
\\ \hline 
         Vascular Lesions (vasc)
& 0.725
\\ \hline 
         Dermatofibroma (df)
& \textbf{0.125}
\\ \hline
 \textbf{Overall Test accuracy}&0.53\\\hline
    \end{tabular}
    
\vspace{-2em}    
\end{table}

\subsubsection{Inference:}
From Table \ref{tab:CNN_Random_Selection} we can observe that accuracy on the Dermatofibroma class has improved from their initial values in Table \ref{tab:Classwise_CNN_preaugmentation}. As we have selected the seed images randomly, we can assume that it is possible to get better results by training the One-Shot GANs on more diverse images which we select on purpose.

\subsection{Subset Selection and One-Shot GANs} \label{sec:Subset_OneShot}

To boost accuracies, we utilize subset selection method \cite{Submodlib} to select the
most diverse images of the Dermatofibroma class present in the training
set. After these images have been selected the same procedure is followed as done in Section \ref{sec:Random_OneShot} to select the best generated images.\\
\vspace{-2.5em}
\subsubsection{Results:} Similar to section \ref{sec:Random_OneShot}, the accuracy of the CNN was found post
augmentation of the training dataset with 100 samples generated by One-Shot GANs.

\begin{table}
    \centering
\vspace{-2em}    
\caption{ Class wise accuracies of CNN by Subset Selection}
\label{tab:CNN_Classwise_Subset}
    \begin{tabular}{|c|c|} \hline 
         \textbf{Class Name}
&   \textbf{Accuracy}
\\ \hline 
         Melanocytic nevi (nv)
& 0.925
\\ \hline 
         Melanoma (mel)
& 0.425
\\ \hline 
         Benign Keratosis-like lesions (bkl)
& 0.60
\\ \hline 
         Basal Cell carcinoma (bcc)
& 0.55
\\ \hline 
         Actinic Keratoses (akiec)
& 0.425
\\ \hline 
         Vascular Lesions (vasc)
& 0.775
\\ \hline 
         Dermatofibroma (df)
& \textbf{0.25}
\\ \hline
 \textbf{Overall Test accuracy}&0.55\\\hline
    \end{tabular}
    
\vspace{-3.5em}    
\end{table}


\subsubsection{Inference:}
We see a significant improvement in accuracies in Table \ref{tab:CNN_Classwise_Subset}compared to Table \ref{tab:Classwise_CNN_preaugmentation} and \ref{tab:CNN_Random_Selection}. This proves that subset selected images provide much better seed for augmentation using One-Shot GANs compared to the random selection method. To assess if the accuracy can be boosted further, we assess the validity of the FID score as a method for selecting generated images.
\vspace{-1.5em}
\subsection{FID Score Assessment to optimize generated image selection}

After exploring the effectiveness of Subset Selection of images we move
forward and assess the importance of selected image being similar to the
original image. In previous experiments in Sections \ref{sec:Random_OneShot} and \ref{sec:Subset_OneShot} we have
utilized the images with the lowest FID score. In this section we
explore the possibility that the lowest FID score might not produce the
highest accuracy on minority class.

\vspace{-1.5em}

\subsubsection{Results:}
For this assessment FID scores have been utilized. Three Categories are
defined for this experiment, they are Bottom (lowest FID scores), Top
(highest FID score) and Random (random selection). The Generated images
are selected to fulfil these three categories and their class-wise and
overall accuracies are displayed in Tables \ref{tab:Classwise_CNN_Similarity} below.

\begin{table}
    \centering
\vspace{-2em}    
\caption{Class wise accuracies of CNN by Similarity}
\label{tab:Classwise_CNN_Similarity}
    \begin{tabular}{|c|c|l|l|} \hline 
         \textbf{Class Name}
&  \textbf{Bottom Selection}
& \textbf{Top Selection}
&\textbf{Random Selection}
\\ \hline 
         Melanocytic nevi (nv)
&  0.925
&         0.95&0.95
\\ \hline 
         Melanoma (mel)
&  0.375
& 0.45
&0.375
\\ \hline 
         Benign Keratosis-like lesions (bkl)
&  0.60
& 0.475
&0.40
\\ \hline 
         Basal Cell carcinoma (bcc)
&  0.55
& 0.625
&0.65
\\ \hline 
         Actinic Keratoses (akiec)
&  0.425
& 0.55
&0.45
\\ \hline 
         Vascular Lesions (vasc)
&  0.775
& 0.825
&0.825
\\ \hline 
         Dermatofibroma (df)
&  \textbf{0.25}
& 0.075
&     0.05
\\ \hline
 \textbf{Overall Test accuracy}& 0.55& 0.56&0.52\\\hline
    \end{tabular}
    
\vspace{-1.5em}    
\end{table}


\subsubsection{Inference:}
It can be noted from Table \ref{tab:Classwise_CNN_Similarity} that the class wise accuracy for the
minority class Dermatofibroma for the Bottom Selection method is higher
than Top Selection method. We can conclude that selection using the
lowest FID score is the most optimal method. It is also important to
note that the FID metric is frequently used for GANs models that are
trained on a diverse training dataset. One-Shot GANs generate very
similar images from a single sample to augment a minority class. This
could potentially increase the risk of overfitting while training a
classifier on the given dataset. To find out if another selection method is
effective, a new metric was designed and tested out.

\vspace{-1em}
\subsection{Content-Space Assessment to optimize generated image selection} \label{sec:New_Metric_Assessment}
We assigned different weights to both the Content and Spatial Branches of the metric. Referencing Equation \ref{eqn_Context_Space} we assigned the weights following values w1=1 and w2=0, w2=1 and w1=0 and w1=w2=0.5.
We can name the conditions as Context only, Space Only and Equal weight respectively.
\vspace{-1.75em}
\subsubsection{Results:}
Following the calculations utilizing the weights the
images with the lowest scores (most similar) were taken for each of the
three categories.

\begin{table}
    \centering
\vspace{-2em}    
\caption{Class wise accuracies of CNN by different weights}
\label{tab:Classwise_CNN_new metric}
    \begin{tabular}{|c|c|c|c|} \hline 
         \textbf{Class Name}
&  \textbf{Content only}&  \textbf{Equal weights}
& \textbf{Space only}
\\ \hline 
         Melanocytic nevi (nv)
&  0.925
&  0.95
& 0.925
\\ \hline 
         Melanoma (mel)
&  0.40
&  0.40
& 0.35
\\ \hline 
         Benign Keratosis-like lesions (bkl)
&  0.525
&  0.50
& 0.55
\\ \hline 
         Basal Cell carcinoma (bcc)
&  0.675
&  0.65
& 0.60
\\ \hline 
         Actinic Keratoses (akiec)
&  0.40
&  0.475
& 0.475
\\ \hline 
         Vascular Lesions (vasc)
&  0.75
&  0.825
& 0.825
\\ \hline 
         Dermatofibroma (df)
&  0.075
&  0.10
&      \textbf{0.375}
\\ \hline
 \textbf{Overall Test accuracy}& 0.53& 0.55&0.58\\\hline
    \end{tabular}   
\end{table}
\vspace{-5mm}
\subsubsection{Inference:}
Following this we can conclude that utilizing Space Only selection is
the best method. It provides the highest accuracy on the CNN post
augmentation of the dataset. We utilize this method of selection and
compare its results with those of FID based selection.
\vspace{-0.5em}

\subsection{  Metric Comparison}
In this subsection we compare the effects of using the two different
metric, FID value and the Context Space assessment metric on the
accuracy of each individual class and the overall accuracy of the model.
For this we have used the Space only mode Content Space metric as seen in Table \ref{tab:Classwise_CNN_new metric} and compared it with values obtained via FID scores from the column labelled 'Bottom' of Table \ref{tab:Classwise_CNN_Similarity}.

\begin{table}
    \centering
\vspace{-2em}    
\caption{Class wise accuracies of CNN by different metrics}
\label{tab:Classwise_by_metric}
    \begin{tabular}{|c|c|c|} \hline 
         \textbf{Class Name}
&  \textbf{Context-Space Metric}
& \textbf{FID Score}
\\ \hline 
         Melanocytic nevi (nv)
&  0.925
& 0.925
\\ \hline 
         Melanoma (mel)
&  0.35
& 0.375
\\ \hline 
         Benign Keratosis-like lesions (bkl)
&  0.55
& 0.60
\\ \hline 
         Basal Cell carcinoma (bcc)
&  0.60
& 0.55
\\ \hline 
         Actinic Keratoses (akiec)
&  0.475
& 0.425
\\ \hline 
         Vascular Lesions (vasc)
&  0.825
& 0.775
\\ \hline 
         Dermatofibroma (df)
& \textbf{ 0.375}
& 0.25
\\ \hline
 \textbf{Overall Test accuracy}& 0.58&0.55\\\hline
    \end{tabular}
    
\vspace{-1.5em}   
\end{table}


\vspace{-0.5cm}
\subsubsection{Inference:}
We can see that the Context Space Assessment gives better results for a
CNN than FID score selection. This shows us that this is the ideal
method of selection for generated images. The best possible model for
one-shot involves Subset-Selected seed images and Context-Space Metric
selected generated images which will be used to augment the training
set. 
\vspace{-1em}
\subsection{Baseline comparison utilizing WGANs and FID model}
To display the effectiveness of the Content-Space Assessment metric and One-Shot-GANs model we can compare the results that this solution gives with baseline data. For the baseline data, we utilize two of the most common methods instead of One-Shot-GANs and context-space metric. The baseline utilizes a WGANs model which generates 1000 images after being trained on the same 10 Subset Selected images from the Dermatofibroma class for 150000 epochs. Following this we select the 100 most similar images using FID score as the selection criteria.  

\begin{table}
    \centering
\vspace{-2em}    
\caption{Classwise accuracy comparison with baseline model}
\label{tab:Baseline}
    \begin{tabular}{|c|c|c|} \hline 
         Class Name &  One-Shot and Content-Space& WGANs-FID\\ \hline 
         Melanocytic nevi (nv)
&  0.925& 0.95\\ \hline 
         Melanoma (mel)
&  0.35& 0.375\\ \hline 
         Benign Keratosis-like lesions (bkl)
&  0.55& 0.575\\ \hline 
         Basal Cell carcinoma (bcc)
&  0.60& 0.575\\ \hline 
         Actinic Keratoses (akiec)
&   0.475& 0.475\\ \hline 
         Vascular Lesions (vasc)
&   0.825& 0.775 \\ \hline 
         Dermatofibroma (df)
&   \textbf{0.375} & 0.05\\ \hline
 \textbf{Overall Test accuracy}& 0.58&0.53\\\hline
    \end{tabular}
    
\vspace{-3em}   
\end{table}


\subsubsection{Inference:}
From Table \ref{tab:Baseline} we can see that the combination of One-Shot GANs along with Content-Space assessment gives significantly better results than the baseline solution which Combines WGANs and FID score. To be precise there is a 37\% increase in accuracy on the minority class of Dermatofibroma for our proposed solution when compared to baseline solution.
\vspace{-1.5em}
\section{Accuracy on different Classifiers}
\vspace{-1.0em}
For this section, we tried to utilize the One-Shot GANs Content-Space Assessment method for different classifier models. We selected 10 most similar generated images from each of the 10 subset selected seed images. This gave us a total of 100 generated images for augmentation of training set. 
Three different Neural Networks were used, Convolutional Neural Network, ResNet50 and XCeption \cite{kaggle_code}. This was done to observe the effect of more model complexity on accuracy. The results of this experiment can be seen in Table \ref{tab:Classwise_Classifiers} below. 
\vspace{-1.5em}

\subsubsection{Inference:}
We can see that the performance of more complex and sophisticated models is better when utilizing One-Shot GANs as a solution to long tail problems. This implies the complex models are better to use as classifiers than simple CNN model which is utilized. 

\begin{table}[h]
    \centering
\vspace{-7mm}    
\caption{Classwise Accuracies by Calssifier model utilized}
\label{tab:Classwise_Classifiers}
    \begin{tabular}{|c|c|c|c|} \hline 
         Class Name&  CNN&  ResNet50& XCeption\\ \hline 
         Melanocytic nevi (nv)
&  0.925&  0.875& 0.90\\ \hline 
         Melanoma (mel)
&  0.35&  0.575& 0.50\\ \hline 
         Benign Keratosis-like lesions (bkl)
&  0.55&  0.625& 0.575\\ \hline 
         Basal Cell carcinoma (bcc)
&  0.60&  0.625& 0.80\\ \hline 
         Actinic Keratoses (akiec)
&  0.475&  0.70& 0.55\\ \hline 
         Vascular Lesions (vasc)
&  0.825&  0.675& 0.825\\ \hline 
         Dermatofibroma (df)&  0.375&  0.375& \textbf{0.575}\\ \hline 
 Test Accuracy& 0.585714& 0.635714&\textbf{0.675000}\\ \hline
    \end{tabular}
    
\vspace{-3.5em}   
\end{table}

\section{ Limitations and Future Directions}
\vspace{-1.5mm}

With our current model the class accuracy on minority class of Dermatofibroma increases significantly from 0\% to 37.5\%, but if we observe Tables \ref{tab:Classwise_CNN_preaugmentation} through \ref{tab:Classwise_by_metric} we can observe that the overall accuracy of the model on the test set does not change. If we compare the results of the best possible methods (on CNN) in Table \ref{tab:Classwise_by_metric} with the initial results in Table \ref{tab:Classwise_CNN_preaugmentation}, we observe that the overall accuracy only increases by 4\%. This shows the limitation of this model and metric when it comes to improving overall accuracy on all the classes. This limitation is possibly caused by the addition of other samples skewing the dataset more in favor of the Dermatofibroma class. To solve this issue future explorations on this topic should deal with the problem of rebalancing the training dataset effectively such that the overall accuracy gets boosted significantly along with accuracy on minority classes. 

%
%
%
\vspace{-3mm}

\bibliographystyle{splncs04}
\bibliography{references}

\begin{thebibliography}{10}
\providecommand{\url}[1]{\texttt{#1}}
\providecommand{\urlprefix}{URL }
\providecommand{\doi}[1]{https://doi.org/#1}

\bibitem{auto_segment}
Al-abayechi, A.A.A., Guo, X., Tan, W.H., Jalab, H.A.: Automatic skin lesion segmentation with optimal colour channel from dermoscopic images. Science Asia  \textbf{40(Suppl. 1)}, ~1--7 (2014). \doi{: 10.2306/scienceasia1513-1874.2014.40S.001}, \url{https://www.scienceasia.org/content/viewabstract.php?ms=4929&v=27&abst=1}

\bibitem{kaggle_code}
de~Arajuo, J.N.: Ham10000: Analysis and model comparison. Kaggle, \url{https://www.kaggle.com/code/jnegrini/ham10000-analysis-and-model-comparison}

\bibitem{Bhattacharya_distance}
Bhattacharyya, A.: On a measure of divergence between two multinomial populations. Sankhyā: The Indian Journal of Statistics (1933-1960)  \textbf{7}(4),  401--406 (1946)

\bibitem{GANs_Paper}
Goodfellow, I., Pouget-Abadie, J., Mirza, M., Xu, B., Warde-Farley, D., Ozair, S., Courville, A., Bengio, Y.: Generative adversarial nets. In: Ghahramani, Z., Welling, M., Cortes, C., Lawrence, N., Weinberger, K. (eds.) Advances in Neural Information Processing Systems. vol.~27. Curran Associates, Inc. (2014), \url{https://proceedings.neurips.cc/paper_files/paper/2014/file/5ca3e9b122f61f8f06494c97b1afccf3-Paper.pdf}

\bibitem{FID_paper}
Heusel, M., Ramsauer, H., Unterthiner, T., Nessler, B., Hochreiter, S.: Gans trained by a two time-scale update rule converge to a local nash equilibrium. In: Guyon, I., Luxburg, U.V., Bengio, S., Wallach, H., Fergus, R., Vishwanathan, S., Garnett, R. (eds.) Advances in Neural Information Processing Systems. vol.~30. Curran Associates, Inc. (2017), \url{https://proceedings.neurips.cc/paper_files/paper/2017/file/8a1d694707eb0fefe65871369074926d-Paper.pdf}

\bibitem{Bhattacharya_distance_signal}
Kailath, T.: The divergence and bhattacharyya distance measures in signal selection. IEEE Transactions on Communication Technology  \textbf{15}(1),  52--60 (1967). \doi{10.1109/TCOM.1967.1089532}

\bibitem{Submodlib}
Kaushal, V., Ramakrishnan, G., Iyer, R.: Submodlib: A submodular optimization library. arXiv  (2022), \url{https://arxiv.org/abs/2202.10680}

\bibitem{Deep_Learning_Mammogram}
Kooi, T., Litjens, G., {van Ginneken}, B., Gubern-Mérida, A., Sánchez, C.I., Mann, R., {den Heeten}, A., Karssemeijer, N.: Large scale deep learning for computer aided detection of mammographic lesions. Medical Image Analysis  \textbf{35},  303--312 (2017). \doi{https://doi.org/10.1016/j.media.2016.07.007}, \url{https://www.sciencedirect.com/science/article/pii/S1361841516301244}

\bibitem{Deep_Learning_Medicine}
Lee, J.G., Jun, S., Cho, Y., Lee, H., Kim, G.B., Seo, J.B., Kim, N.: Deep learning in medical imaging: General overview. Korean Journal of Radiology  \textbf{18},  570 -- 584 (2017), \url{https://api.semanticscholar.org/CorpusID:4345827}

\bibitem{WGANs}
Martin~Arjovsky, Soumith~Chintala, L.B.: Wasserstein generative adversarial networks. In: Volume 70: International Conference on Machine Learning. pp. 214--223. Proceedings of Machine Learning Research, Sydney, NSW, Australia (2017), \url{https://proceedings.mlr.press/v70/arjovsky17a.html}

\bibitem{otsu}
Otsu, N.: A threshold selection method from gray-level histograms. IEEE Transactions on Systems, Man, and Cybernetics  \textbf{9}(1),  62--66 (1979). \doi{10.1109/TSMC.1979.4310076}

\bibitem{One_Shot_GANs}
Sushko, V., Gall, J., Khoreva, A.: Learning to generate novel scene compositions from single images and videos. arXiv  (2021), \url{https://arxiv.org/abs/2105.05847}

\bibitem{HAM10000_dataset}
Tschandl, P., Rosendahl, C., Kittler, H.: The ham10000 dataset, a large collection of multi-source dermatoscopic images of common pigmented skin lesions. Scientific Data  (2018). \doi{https://doi.org/10.1038/sdata.2018.161}

\bibitem{Otsu_kmeans}
Yu, D.L.J.: Otsu method and k-means. In: Ninth International Conference on Hybrid Intelligent Systems. IEEE, Shenyang, China (2009). \doi{10.1109/HIS.2009.74}

\bibitem{Long_tail_survey}
Zhang, Y., Kang, B., Hooi, B., Yan, S., Feng, J.: Deep long-tailed learning: A survey. IEEE Transactions on Pattern Analysis and Machine Intelligence  \textbf{45}(9),  10795--10816 (2023). \doi{10.1109/TPAMI.2023.3268118}

\end{thebibliography}
\vspace{-2em}


\end{document}